\documentclass[12pt,preprint]{aastex}

\usepackage{lscape}
\begin{document}


\title{Suzaku Monitoring of the Iron K Emission Line in the Type 1 AGN NGC 5548}


\author{Yuan Liu\altaffilmark{1,2},
Martin Elvis\altaffilmark{1}, Ian M. McHardy\altaffilmark{3}, Dirk
Grupe\altaffilmark{4}, Belinda J. Wilkes\altaffilmark{1},
 James Reeves\altaffilmark{5}, Nancy Brickhouse\altaffilmark{1}, Yair Krongold\altaffilmark{6},
 Smita Mathur\altaffilmark{7}, Takeo Minezaki\altaffilmark{8}, Fabrizio Nicastro\altaffilmark{1},
 Yuzuru Yoshii\altaffilmark{8}, Shuang Nan
Zhang\altaffilmark{2,9,10}}

\altaffiltext{1}{Harvard-Smithsonian Center for Astrophysics, 60
Garden Street, Cambridge, MA 02138, USA; yliu@cfa.harvard.edu,
elvis@cfa.harvard.edu}
 \altaffiltext{2}{Physics
Department and Center for Astrophysics, Tsinghua University, Beijing
100084, China; yuan-liu@mails.tsinghua.edu.cn}
 \altaffiltext{3}{School of Physics and
Astronomy, University of Southampton, Southampton SO17 1BJ, UK}
\altaffiltext{4}{Department of Astronomy and Astrophysics,
Pennsylvania State University, 525 Davey Lab, University Park, PA 16
802, USA} \altaffiltext{5} {Astrophysics Group, School of Physical
and Geographical Sciences, Keele University, Keele, Staffordshire
ST5 5BG, UK} \altaffiltext{6}{Instituto de Astronomia, Universidad
Nacional Autonoma de Mexico, Apartado Postal 70-264, 04510 Mexico
DF, Mexico} \altaffiltext{7}{Department of Astronomy, The Ohio State
University, 140 West 18th Avenue, Columbus, OH 43 210, USA}
\altaffiltext{8}{Institute of Astronomy, School of Science,
University of Tokyo, Mitaka, Tokyo 181-0015, Japan}

\altaffiltext{9}{Key Laboratory of Particle Astrophysics, Institute
of High Energy Physics, Chinese Academy of Sciences, P.O.Box 918-3,
Beijing 100049, China} \altaffiltext{10}{Physics Department,
University of Alabama in Huntsville, Huntsville, AL 35899, USA}



\begin{abstract}
We present 7 sequential weekly  observations of NGC 5548 conducted
in 2007 with the \textit{Suzaku} X-ray Imaging Spectrometer (XIS) in
the 0.2-12 keV band and Hard X-ray Detector (HXD) in 10-600 keV
band. The iron K$\alpha$ line is well detected in all seven
observations and K$\beta$ line is also detected in four
observations. In this paper, we investigate the origin of the Fe K
lines using both the width of the line and the reverberation mapping
method.

 With the co-added XIS and HXD spectra, we identify
Fe K$\alpha$  and K$\beta$ line at 6.396$_{-0.007}^{+0.009}$ keV and
7.08$_{-0.05}^{+0.05}$ keV, respectively. The width of line obtained
from the co-added spectra is 38$_{-18}^{+16}$ eV
($\textrm{FWHM}=4200_{-2000}^{+1800}$ km/s) which corresponds to a
radius of 20$_{-10}^{+50}$ light days, for the virial production of
 $1.220\times10^7$ M$_{\odot}$ in NCG 5548.

To quantitatively investigate the origin of the narrow Fe line by
the reverberation mapping method, we compare the observed light
curves of Fe K$\alpha$ line with the predicted ones, which are
obtained by convolving the continuum light curve  with the transfer
functions in a thin shell and an inclined disk. The best-fit result
is given by  the disk case with $i=30^\circ$  which is better than a
fit to a constant flux of the Fe K line at the 92.7\% level
(F-test). However, the results with other geometries are also
acceptable (P$>$50\%). We find that the emitting radius obtained
from the light curve is 25-37 light days, which is consistent with
the radius derived from the Fe K line width. Combining the results
of the line width and variation, the most likely site for the origin
of the narrow iron lines is 20-40 light days away from the central
engine, though other possibilities are not completely ruled out.
This radius is larger than the H$\beta$ emitting parts of the broad
line region at 6-10 light days (obtained by the simultaneous optical
 observation), and smaller than the inner radius of the hot dust in NGC 5548 (at
about 50 light days).
\end{abstract}



\section{Introduction}\label{sec1}

The neutral, photoionized iron K emission line commonly found in the
X-ray spectra of active galactic nuclei (AGNs) can be decomposed
into a narrow core around 6.4 keV and a broad redshifted component,
though sometimes only one of them is present (Fabian et al. 2000;
Yaqoob et al. 2001; Nandra et al. 1997, 2007). The broad,
relativistic component could be modelled as being emitted from the
vicinity of the central black hole and used to determine the
parameters of the black hole (e.g. Brenneman \& Reynolds 2006). An
alternative explanation of this feature is the complex absorption
origin (Miller, Turner \& Reeves 2008). However, the origin of the
narrow component is still unclear, but could be critical to
understanding the central structure of AGNs. The narrow width (FWHM
$\sim$ several thousand km/s or less, e.g. Yaqoob \& Padmanabhan
2004) suggests an origin well outside the continuum producing
region. One of the proposed sites is the putative `obscuring torus'
(e.g. Nandra 2006) at $\sim$0.1-1 pc. The evaporation radius of the
dust can be estimated by
$r=1.3L^{1/2}_{\textmd{uv},46}T^{-2.8}_{1500}$ pc (Barvainis 1987),
where $L_{\textmd{uv},46}$ is the ultraviolet luminosity in units of
10$^{46}$ ergs/s, and $T_{1500}$ is the evaporation temperature in
units of 1500 K. For NGC 5548, since $L_{\textmd{uv},46}\sim0.01$
and assuming $T_{1500}=1$, the evaporation radius is about 0.1 pc.
However, the broad emission lines have similar line widths, so the
more compact broad line region (BLR) is another possible location
which cannot be ruled out simply. A BLR origin is supported by the
quasi-simultaneous optical spectroscopic observation with
\textit{Chandra} observation of NGC 7213 (Bianchi et al. 2008),
which shows consistent Fe K and H$\beta$ line widths, and by the
rapid $N_H$ changes seen in several AGNs (Elvis et al. 2004;
Puccetti et al. 2007; Risaliti et al. 2002), which require a
BLR-like radius for the $N_H\gtrsim10^{23}$ cm$^{-2}$ absorbers.
Seen from another angle, these absorbers must re-emit in Fe K.

The lag between the variation of the flux of the continuum and the
line can be used to measure the location of an emission line region
(Blandford \& McKee 1982) and this `reverberation mapping'
methodology has been applied to the optical and UV broad emission
lines (BELs) with great success (Peterson et al. 2004). However,
until now the 10\% or greater error on the flux of the 6.4 keV Fe K
line (compared with the usual 1\%-5\% error on the flux of BELs),
and the low sampling frequency of the X-ray observations, have made
it hard to apply this method to determine lags for Fe K lines,
especially on relatively short timescale expected ($\sim$10 days for
a BLR origin). For example, Chiang et al. (2000) found the flux of
the iron K line in NGC 5548 was consistent with being constant using
the four simultaneous observations of \textit{ASCA} and
\textit{RXTE} within 25 days and another observation after about
half a year. However, some response of the Fe K line to the
continuum changes has been found. Markowitz et al. (2003) tried this
method, analyzing the long-term \textit{RXTE} spectra for seven
sources. Although they found no evidence for correlated variability
between the line and continuum, comparable systematic long-term
($\sim$3-4 years) decreases in the line and continuum were present
in NGC 5548.

We report here on a series of seven sequential X-ray observations of
NGC 5548 by \textit{Suzaku}, spaced roughly weekly. This observing
campaign was designed, among other goals, to constrain the flux of
the  iron line to about 10\% in each observation. This allows us,
for the first time, to apply the reverberation mapping technique to
Fe K to try to distinguish different geometries of the Fe K emitting
region. NGC 5548 is the source best studied by the optical
reverberation mapping technique (e.g. Peterson \& Wandel 1999) and
so has the best determined radial structure. Therefore, it is easy
to determine the relative location of the Fe K line emitting region.
In previous observations, only the Fe K$\alpha$ line was detected.
As we will present in this paper, the K$\beta$ line is also well
detected in the \textit{Suzaku} observations. Although very weak in
some observations, K$\beta$ is useful to constrain the ionization
state of iron.

The iron line in NGG 5548 has been observed by several X-ray
satellites: \textit{ASCA} spectra suggested a relativistic broad
iron K$\alpha$ line in NGC 5548 with $\sigma=340_{-120}^{+190}$ eV
(68\% error for 4 interesting parameters, Nandra et al. 1997) and
$\sigma\sim 400$ eV (Chiang et al. 2000). However, less than two
years later, only the narrow K$\alpha$ line was detected in a
\textit{Chandra} observation with much higher energy resolution (38
eV vs 160 eV) but lower S/N (Yaqoob et al. 2001). An
\textit{XMM-Newton} EPIC CCD observation confirmed the
\textit{Chandra} result (Pounds et al. 2003).

In \S2, we describe the observations and the procedure of the data
reduction. In this paper we intend to investigate the origin of the
Fe K line using the reverberation mapping method and the width of
the line. Therefore, in \S3, we first fit the spectra of each
observation to determine the flux of the continuum and the iron
line. We then fit the co-added spectra in \S3.2 to determine the
mean parameters, especially the width of the iron line. In \S4.2 and
\S4.3, we calculate the transfer functions in different geometries
and investigate the possible emitting region of
 iron line. In \S5.1, we discuss the possible origin of the iron line. In \S5.3, we briefly discuss the implications of the intensity and the equivalent
 width of iron line. In \S6 we give our
 conclusions. Throughout this paper we adopt the redshift of
 NGC 5548 obtained from 21 cm H I measurements, i.e. z=0.017175 (de Vaucouleurs et al. 1991). The errors quoted in this paper correspond to 90\%
 confidence level ($\Delta\chi^2=2.706$, Avni 1976) if not otherwise specified.

\section{Observations and data reduction}\label{sec2}
\subsection{Observations}\label{sec21}

During 2007 June to 2007 August, NGC 5548 was observed by the CCD
X-ray Imaging Spectrometers (XIS 0, 1, and 3) in 0.2-12 keV band
 (Koyama et al. 2007)
 and by the
Hard X-ray Detector (HXD) in 10-600 keV band (Takahashi et al. 2007)
on \textit{Suzaku} (Mitsuda et al. 2007)   seven times for 28.9
ks-38.7 ks each. We denote these as observations 1-7. The details of
the observations are summarized in Table 1.

\subsection{Data reduction}\label{sec22}

Following the standard procedures outlined in the ``Suzaku Data
Reduction (ABC) Guide (version
2)\footnote{http://heasarc.gsfc.nasa.gov/docs/suzaku/analysis/abc/}'',
we used the updated Charge Transfer Inefficiency (CTI) calibration
(Suzaku XIS CALDB 20081110) and screened the events using the
\texttt{xispi} (Ftools 6.5) and \texttt{xselect} scripts
 provided by \textit{Suzaku} team (we adopted the standard criterion in xis\_event.sel and xis\_mkf.sel)\footnote{http://suzaku.gsfc.nasa.gov/docs/suzaku/analysis/xisrepro.xco,
http://suzaku.gsfc.nasa.gov/docs/suzaku/analysis/xis\_event.sel,
http://suzaku.gsfc.nasa.gov/docs/suzaku/analysis/xis\_mkf.sel},
respectively.

X-ray spectra were extracted using
\texttt{xselect}\footnote{http://heasarc.nasa.gov/docs/software/lheasoft/ftools/xselect/index.html}
from all the XISs with a circular extraction region of radius 260
arcsec centered on NGC 5548 ($\alpha=$14h 17m 59.5s, $\delta=$+25d
08m 12.4s, J2000). Background spectra were obtained from a larger
annulus around the source (but avoiding the calibration sources on
the corners of the chips). Response matrices (\texttt{rmf}) and
effective area (\texttt{arf}) files were generated with the
\texttt{xisrmfgen}\footnote{http://heasarc.nasa.gov/docs/suzaku/analysis/xisrmfgen.html}
and
\texttt{xissimarfgen}\footnote{http://heasarc.gsfc.nasa.gov/docs/suzaku/analysis/xissimarfgen/}
(\texttt{estepfile=dense} and \texttt{num\_photon=300000}),
respectively. We then combined the spectra, background,
\texttt{rmf}, and \texttt{arf} files from the two front-side
illuminated CCDs, XIS 0 and 3, for each observation
 with \texttt{addascaspec}\footnote{http://heasarc.gsfc.nasa.gov/lheasoft/ftools/fhelp/addascaspec.txt} (we
 denote the combined spectra by `XIS03').
  The spectra of XIS 1 are considered separately, as this detector uses a back-side illuminated CCDs.

NGC 5548 was detected by the silicon diode PIN instrument of the
HXD, but below the sensitivity limit of the GSO crystal scintillator
instrument. We downloaded the tuned non-X-ray background (NXB,
version 2.0) files from the \emph{Suzaku} Guest Observer Facility
(GOF)\footnote{ftp://legacy.gsfc.nasa.gov/suzaku/data/background/pinnxb\_ver2.0\_tuned/}
and then merged the good-time interval (GTI) of the NXB with that of
the screened event files to produce a common GTI using
\texttt{mgtime}. Then the spectra of the source and the NXB were
extracted by \texttt{xselect} using the common GTI. And the dead
time of the source spectra was corrected by \texttt{hxddtcor}. Since
the event rate in the PIN background event file is 10 times higher
than the real background to suppress the Poisson errors, the
exposure time of the spectra of NXB was increased by a factor of 10.
The cosmic X-ray background (CXB) was not taken into account in the
NXB file, therefore we also added a CXB component in the spectral
fitting using the model given by the ABC guide (Section 7.3.3). The
response file \texttt{ae\_hxd\_pinxinome3\_20080129.rsp} was used
for observation 1-5, while the response file
\texttt{ae\_hxd\_pinxinome4\_20080129.rsp} was used for observation
6 and 7 due to the changes in instrumental settings of
\textit{Suzaku} during different observation epochs. The PIN
 spectra were multiplied by a  constant
cross-normalization factor of 1.16 to account for the differences in
calibration between XIS
 and PIN\footnote{http://heasarc.gsfc.nasa.gov/docs/suzaku/analysis/watchout.html}.

\section{Spectral fitting}\label{sec3}

To perform the reverberation mapping calculation, we should first
determine the flux of the continuum and the Fe K line. In order to
avoid the influence of the complex absorption below 3 keV (Detmers
et al. 2008; Steenbrugge et al. 2003), we only analyze the XIS
spectra  in 3-10 keV band. The warm absorber seen in the $<$3 keV
spectra is presented by Krongold et al. (2009). In this paper, we
will focus on the property and origin of the iron emission line. A
global fit to the entire \textit{Suzaku}  spectral band for the
entire campaign will be presented in a subsequent paper.

 For each observation, we fitted the spectra of XIS03 simultaneously with
the spectra of XIS 1 in XSPEC (version 12.4). Although both of the
energy resolution and the effective area of XIS 1 are lower than XIS
0, 3 in 4-10 keV band (the background level of XIS 1 is also higher
than XIS 0, 3 in this band), it is still useful to reduce the error
of the intensity of the Fe K line. We fixed the Galactic absorption
column at 1.63$\times10^{20}$ cm$^{-2}$ (Murphy et al. 1996), and
fitted the spectra with a single power law. The continua were well
described by a single power law except for the region of the iron
K$\alpha$ and K$\beta$ lines around 6.4 keV and 7.0 keV,
respectively (see Figure 1). Therefore, we added two Gaussian lines
to describe these features. It is important to construct a
self-consistent model to describe all the components of the
continuum and theoretically predict the strength of the Fe K line,
as in Murphy \& Yaqoob (2009). However, this is not the purpose of
this paper. Our purpose is only to simply model the Fe K line using
a Gaussian line and determine the flux and width of it and then to
perform the reverberation calculation. A K$\alpha$ line is required
by all the seven spectra ($>$6 $\sigma$), while K$\beta$ line is
only required by four observations ($>$2 $\sigma$ for 1, 2, 4, and
7). The 90\% upper limit of the flux of the K$\beta$ line is
determined in observation 3, 5, and 6. Since the K$\beta$ line is
weak, we fixed the width of K$\beta$ to be the same as that of
K$\alpha$. Fe K$\beta$ line was also detected in other sources, e.g.
NGC 2992 (Yaqoob et al. 2007) and Mrk 3 (Awaki et al. 2008).
 The detailed result of the fitting is given in Table 3, though we are mainly concerned about the
 flux of the continuum and the intensity of the Fe K line. Due to the weak reflection
 component in NCG 5548 (e.g. Pounds et al. 2003), if we simultaneously
 fit the continua in the XIS and PIN spectra of each observation using the \texttt{pexrav} model instead of
 the power law model, the intensity of the Fe K line
 will systematically decrease by 10\% for each observation (the strength of the reflection component $R \sim 0.2-1.5$). However, this
 change will not influence the reverberation mapping result in \S4.3, since it degenerates with the normalization of
 the transfer function (see the detail of the transfer function in
 \S4.2). The PIN data is also not helpful to reduce the error of the flux of the Fe K line due to the
 additional uncertainty introduced by the reflection component and
 large error in the PIN spectra.
 The detailed discussion about the variation of the X-ray continuum and simultaneous UV/optical data will be presented in
another paper.

\subsection{Continuum light curve}\label{sec31n}
Having fitted the spectra, we calculated the observed flux of the
continuum in the 3-10 keV band. The result light curve is shown in
Figure 3a. To better sample the continuum light curve, we also
utilized one observation of the continuum from the simultaneous
 \textit{Swift} campaign. Since the observation times and the flux of other
\textit{Swift} data are quite similar to that of the \textit{Suzaku}
data, we will not include them in this paper. The details of the
\textit{Swift} campaign will be discussed in Grupe et al. (2009). 28
observations of the continuum with PCA on \textit{RXTE} before the
\textit{Suzaku} campaign are also used, since we are looking for the
lag between the variation of continuum and line. However, due to the
short exposure time, we cannot determine the flux of the line in any
of these additional observations. The details of the \textit{RXTE}
and \textit{Swift} observations are summarized in Table 2.

\subsection{Narrow iron lines}\label{sec31}
To determine the mean parameters of the Fe K$\alpha$ and K$\beta$
 lines more accurately, especially the width of the line, we added the spectra from the seven
observations together using \texttt{addascaspec} (the XIS03, XIS1,
and PIN spectra were added separately and then fitted
simultaneously). The net counts of the source in 3-10 keV band in
the total XIS03 and XIS1 spectra are 160349 and 75332, respectively.
The net counts of the source in 12-35 keV band in the total PIN
spectra is 22243. We found that the XIS1 spectra could not provide
any useful constraint on the width of the Fe K line, since the
$\sigma$ of the Gaussian line is pegged at 0 eV (the 90\% upper
limit is 30 eV). Therefore, we will only utilize the co-added XIS 03
and PIN spectra to determine the width of the Fe K line.

If we only fitted the co-add XIS03 spectra using the model in \S3,
i.e. a power law and two Gaussian lines, the width is
$\sigma=50_{-15}^{+14}$ eV.  However, since the weak reflection
component could influence the width of the Fe K line, we then
simultaneously fitted the co-added XIS 03 and PIN spectra using the
\texttt{pexrav} model and two Gaussian lines. The derived parameters
are  given in Table 4, where the strength of the reflection
component ($R\sim0.8$) could explain the Fe K line. As shown in
Figure 1, only a narrow iron line is clearly present in the spectra,
with a width, $\sigma=$38$_{-18}^{+16}$ eV. This value is consistent
with previous results: i.e. $\sigma=$41$_{-24}^{+32}$ eV obtained by
HEG+MEG on \textit{Chandra} (Yaqoob et al. 2001) and
$\sigma=$40$_{-40}^{+40}$ eV (MOS) and 64$_{-24}^{+24}$ eV (PN)
obtained by \textit{XMM-Newton} (Pounds et al. 2003).

Since the peak energies of the line and the parameters of the
continua are somewhat different for each observation, we
simultaneously fitted the spectra of all observations in XSPEC to
test whether the line was broadened artificially in the co-added
process. We required the width of line to be the same for all
observations and kept other parameters free. The obtained width is
only slightly smaller than that from the co-add spectra by about 2
eV, which implies the co-added method has not significantly
broadened the width.

The width obtained by the co-add
 spectra is
 inconsistent with zero at 2.2 $\sigma$ and corresponds to FWHM= 4200 km/s and a radius of
5.2$\times10^{16}$ cm or 5.2$\times10^{3}$ $r_g$
  ($r_g=GM/c^2$), for the
 6.71$\times$10$^7 M_{\odot}$  black hole in NGC 5548 (Peterson et al. 2004).
 In \S5.1,
we will estimate the location of the emitting material of the line
and discuss the origin of the line.

To test the presence of the broad, relativistic Fe K$\alpha$ line
found by \textit{ASCA}, we tried the \texttt{diskline} model in
XSPEC to fit the K$\alpha$ line. We fixed the inner and outer disk
radii at 6$r_g$ and 1000$r_g$ ($r_g=GM/c^2$), respectively.  The
index of the power law emissivity was frozen at -2.5 (the mean value
obtained in Nandra et al. 1997). If we thaw the index, it will be
pegged at the positive upper limit in XSPEC, i.e., the flux of the
line is dominated by the outer disk and therefore it is not a disk
line at all. It was found that the \texttt{diskline} model cannot
describe the K$\alpha$ line alone, since $\chi^2$ was higher by 128
than for a narrow line with the same number of parameters. The same
conclusion was also obtained by Yaqoob et al. (2001).

Next, we investigated the result if we fit the K$\alpha$ line with a
disk line and a Gaussian line. Besides the constraint on the
\texttt{diskline} model mentioned above, we also fixed the value of
the inclination angle at 31 degrees, which is the best-constrained
value from the \textit{ASCA} data (Yaqoob et al. 2001). We found the
central value of the intensity of the disk line is pegged at 0 and
the 90\% upper limit is $4\times10^{-6} $ photons/cm$^2$/s.
Therefore, any broad component must be $>$ 5 times weaker than the
narrow component and we will not include this component in the
following discussion.

We show the confidence regions for the peak energies of K$\alpha$
and K$\beta$ in Figure 2a. The expected values of different iron
ionization states (Palmeri et al. 2003; Mendoza et al. 2004; Yaqoob
et al. 2007) are also shown. Since the peak energy may be influenced
by the residuals in the energy scale calibration, we extracted the
spectra of the $^{55}$Fe calibration sources on the corners of the
CCDs and added them together (see the inset in Figure 1b). Using a
Gaussian line to fit the Mn K$\alpha$ line, we found the peak
energies are 5.892$_{-0.001}^{+0.001}$ keV and
5.893$_{-0.001}^{+0.002}$ keV for XIS03 and XIS1, respectively. The
expected value of the Mn K$\alpha$ line is 5.895 keV. This result is
well within the accuracy of the absolute XIS energy scale given by
the \textit{Suzaku} Technical Description
\footnote{http://www.astro.isas.jaxa.jp/suzaku/doc/suzaku\_td/} as
0.2\%\footnote{We found that using the calibration taken for the
same time interval was important. A preliminary analysis using the
calibration available when the observations were made showed a 20 eV
offset, which was puzzling. The amplitude of the Mn K$\alpha$
apparent energy variation was also about 20 eV}. The fit widths of
Mn K$\alpha$ line are 12$_{-5}^{+3}$ eV and 16$_{-8}^{+5}$ eV for
XIS03 and XIS1, respectively. Therefore, the peak energy obtained by
the spectral fitting is reliable and we could conclude that the iron
emission line in NGC 5548 is most likely to be dominated by
relatively low ionization states, $<$Fe XIII (99\% confidence).
Actually, the observed Fe K line could be a blend of several
ionization states.

The confidence regions of the intensities of the Fe K$\alpha$ and
K$\beta$ lines are shown in Figure 2b. The expected ratio of
I(K$\beta$) to I(K$\alpha$) varies from 0.12 to 0.20 for different
ionization states (dashed lines in Figure 2b, Palmeri et al. 2003;
Mendoza et al. 2004;). Our result is fully consistent with the
expected value. However, due to the weakness of K$\beta$, the error
is still too large to constrain the ionization state precisely.

\section{Time variability analysis}\label{sec4n}

\subsection{Fe K$\alpha$ line light curves}\label{sec4}

 The light curve of Fe K$\alpha$ line is shown in Figure 3b each being determined to $\sim$10\%. As K$\beta$ is weak in some
observations, we will only discuss the result for K$\alpha$ below.
As shown in Figure 3a, the flux of the continuum changed strongly
during the seven observations. The highest value
(2.65$\times10^{-11}$ ergs/cm$^2$/s, for observation 5) is about 4
times that of the lowest (6.89$\times10^{-12}$ ergs/cm$^2$/s, for
observation 1). However, perhaps due to the relatively large error,
the flux of the line is consistent with being constant,
$\chi_{min}^2=$3.05 (P=80\%) and the value of the corresponding line
constant flux is 2.24$\times10^{-5}$ photons/cm$^2$/s.

A constant Fe K$\alpha$ flux is the simplest solution to the Fe
K$\alpha$ light curve, and it requires the emitting region should be
far from the central engine to smooth out the variation in short
time scale. According to the result in \S4.3 and Figure 6b, it
should be larger than 100 light days. However, as shown in Figure 1
and 2 in Markowitz et al. (2003), the presence of the variation of
the Fe K flux in time scale shorter than 100 days indicates the
emitting region must be smaller than 100 light days. The emitting
region required by the constant flux fitting is also inconsistent
with that required by the line width (see \S5.1). Therefore, we have
investigated whether there is a better and more self-consistent
solution than the constant flux fitting. To quantitatively access
the emitting location of the Fe K$\alpha$ line, we should perform
the reverberation calculation. The cross-correlation function (CCF)
is usually used in the optical reverberation mapping to determine
the lag between the continuum and the emission line. However, due to
the very few data points of the light curve of the Fe K line  and
the large error of them, we did not find any significant peak in
CCF. Therefore, we will approach this problem in a different way. To
do so we first calculate the transfer function for Fe K$\alpha$ in
\S4.2.

\subsection{Transfer function}\label{sec41}

As in the case of the optical broad emission lines, we calculate a
transfer function for Fe K$\alpha$, i.e. the response of the flux of
the line to a $\delta$ function change in the continuum. We consider
two cases: (1) a spherical region and (2) an inclined disk.

(1) We assume the emitting material is spherically distributed
around the center with an inner ($r_{min}$) and outer ($r_{max}$)
radius. The transfer function in this model is then simply a
constant between 0 and $2r_{min}/c$, and then decays to 0 at
$2r_{max}/c$ (see Figure 4a and Peterson 1993).

(2) For a thin disk, the general form of the transfer function is
two-peaked (Welsh \& Horne 1991). Figure 4b shows the result for
different values of the inclination angle $i$ (the angle between the
normal of the disk and the line of sight).

The shape of the transfer function depends on the form of the
``responsivity" $\varepsilon(r)$ (see Figure 4a), which combines the
effects of the distributions of the number density of clouds and the
emissivity per cloud. We assumed that the responsivity is a power
law $\varepsilon(r)\varpropto r^{-\alpha}$ and the normalization is
adjusted to fit the observed light curve. The power law form is
simple and somewhat arbitrary. However, since we only consider the
thin shell case (i.e. $\Delta r \ll r$, or equivalent to the locally
optimized clouds model), the result is not sensitive to the detailed
form of the responsivity nor the index of the power law (see Figure
6a). $\alpha=3$ is adopted in the calculation in \S4.3.

We could then obtain the predicted light curve of the line by
convolving the transfer function with the light curve of the
continuum flux. In the thin spherical shell and disk cases discussed
in \S4.3, it can be proved that the the lag time, $\tau$, obtained
by the CCF just corresponds to the radius of the emitting region,
i.e. $\tau=r/c$ (Koratkar \& Gaskell 1991).

\subsection{Comparison with the observed light curve}\label{sec42}
Since the observation data of the continuum are still too few to
produce a complete light curve, we performed linear interpolation
between data points to convolve the light curve with the transfer
function.

We used the light curve of the continuum in 3-10 keV band in order
to have precision measurements. Similar bands (e.g. 2-10 keV) have
been adopted in the previous attempts to determine the lag between
the variation of the continuum and the Fe K line (e.g. Markowitz et
al. 2003, 2009), although only photons from above the ionization
threshold can actually lead the emission of an Fe K
photon\footnote{A similar approximation is used in reverberation
mapping of AGNs (e.g. Peterson et al. 2002; Bentz et al. 2007) }. As
shown in Figure 5, the flux in 3-6 keV band is tightly and nearly
proportionally correlated with the flux in 8-10 keV band. Any
possible effects due to rapidly changing $N_H$ (Risaliti et al.
2005) must therefore be small. As a result, the results of the
following calculations will not be sensitive to the adopted energy
band of the continuum. The change of the continuum slope is also
related to the total amount of the ionization flux. However, this is
a minor factor compared with the change of the normalization of the
continuum. Specially, this effect should be even small for NGC 5548,
since e.g. Sobolewska and Papadakis (2009) showed that the continuum
slope is nearly independent on the flux. We find the same lack of
dependence (Table 3).

We fixed the width of the emitting region at 0.1 light days,
substantially smaller than the likely radius of the Fe K emitting
region, and varied $r_{min}$. After convolving the light curve of
the continuum with the transfer functions in different geometries
and $r_{min}$, we compared the predicted light curve of the line
with the observed one to find the minimum value of $\chi^2$ (since
the error on the flux is asymmetric, we conservatively adopt the
larger one to calculate $\chi^2$). Figure 6b shows $\chi^2$ vs
$r_{min}$. All the curves show pronounced minima (somewhat
surprisingly, given the weak structure in the Fe K$\alpha$ light
curve) and finally decrease towards the result of the constant
fitting (the horizontal dashed in Figure 6b) with increasing
$r_{min}$, as the result of the significant smoothing effect with
large $r_{min}$. The values of minimum $\chi^2$ in the spherical
thin shell case and disk case with $i=15^\circ$ and $i=30^\circ$ are
well below that of the constant fitting, and we will only discuss
these three cases below. With the smallest value of $\chi^2$, the
best-fitting light curve in the disk with $i=30^\circ$ is improved
at 92.7\% level compared with the constant fitting. We show the
predicted light curves corresponding to the minima of $\chi^2$ in
these three cases in Figure 7. Due to the few data points, we cannot
tightly constrain the inclination angle. Except for the the disk
case with $i=15^\circ$, the positions of the minimum $\chi^2$ in the
spherical case (27 $_{-7}^{+22}$ light days) and the disk case with
$i=30^\circ$ (37 $_{-5}^{+7}$ light days) are similar, and they are
smaller than the inner radius of the dust (47-53 light days, see the
discussion in \S5.1). The value of the best-fitting $r_{min}$ in the
disk case with $i=15^\circ$ is 61$_{-7}^{+6}$ light days, which is
beyond the inner radius of the dust of NGC 5548 (such optically
thick region will significantly absorb the photons of Fe K lines)
and only marginally consistent with the small tail of the 90\%
confidence interval of the emitting region inferred from the widths
of Fe K$\alpha$ line (see the error bar in Figure 9 and it should be
noted that the error bar is quite asymmetric). In addition,
$i=15^\circ$ is also well smaller than the best-fitting inclination
angle of the disk line in the \textit{ASCA}
 ($i=31.5^\circ\pm6.5^\circ$, Yaqoob et al. 2001), though it is not very reliable, due to the absence of
 the broad component of the Fe K line in the following \textit{Chandra}, \textit{XMM-Newton}, and our \textit{Suzaku} observations. Therefore, the
disk case with $i=15^\circ$ is quite unlikely, and we will only
focus on the spherical case and the disk case with $i=30^\circ$ when
discussing the origin of the narrow Fe K$\alpha$ line in detail in
\S5.1.

\section{Discussion}\label{sec5}

\subsection{Location of the Fe K emitting region}\label{sec51n}
Assuming the geometry and dynamics of  the emitting region of the Fe
K line are the same as that of the H$\beta$ line, and using the
virial relation for the H$\beta$ line, $\sigma^2
r/G=1.220\times10^7$ M$_{\odot}$ (Peterson et al. 2004), the radius
of the Fe K emitting region can be derived using the Fe K$\alpha$
width, $\sigma$, obtained in \S3.2. The derived radius using the
width obtained by the co-added XIS03 and PIN spectra is
20$_{-10}^{+50}$ light days\footnote{Since the virial production is
an observed quantity and  the radius derived from width will be
compared with that also obtained from the reverberation mapping
method, no additional geometry factor is required.}.

In Figure 8, we plot the line width size against $r_{min}$ from the
reverberation analysis (Figure 6b) for both the spherical thin shell
case and the disk case with $i=30^\circ$ with 90\% confidence
intervals for both quantities. The disk case with $i=30^\circ$ and
the spherical case are consistent with the width of the co-added Fe
K$\alpha$ line \footnote{The calibration residual in the width of Mn
K$\alpha$ line derived in \S3.2 (12$_{-5}^{+3}$ eV and
16$_{-8}^{+5}$ eV for XIS03 and XIS1, respectively) is partly due to
the systematic error on the calibration of the non-Gaussian response
function of XIS (Koyama et al. 2007). The energy resolution at the
center of the CCD chip is also slightly better than that at the
corner, but this difference is smaller than the systematic error on
the calibration. Therefore, the true width of the Fe K line could be
smaller than the observed one by a few eVs due to the above factors.
However, these effects could not be  accurately corrected simply.}.
 Combining the above results,
we conclude that the origin of the narrow iron line in NGC 5548 is
likely to be $\sim20-40$ light days away from the continuum source,
for the geometries considered. However, the other possible origins
are not completely ruled out due to the incompleteness of the light
curves, our model-dependent method, and the sizeable error on the
width of the Fe K line.

Since NCG 5548 is one of the best-studied AGNs with reverberation
mapping, the locations of the BELs are well-determined, especially
for the H$\beta$ line. The lag between the flux of H$\beta$ line and
the continuum varies from 6.5 light days to 26.5 light day depending
on the luminosity of the continuum (Bentz et al. 2007; Cackett \&
Horne 2006). Since the correlation between the lag time and the
continuum flux is more significant than the correlation between the
lag time and the width of H$\beta$ line (Bentz et al. 2007), and the
broad component of the  H$\beta$ line is weak during the
\textit{Suzaku} campaign, we will utilize the continuum flux to
predict the radius of the H$\beta$ line region.

From the simultaneous optical spectra of NGC 5548 from FLWO FAST
spectrograph (June 19-23, 2007), we measured the monochromatic flux
at 5100 {\AA} and used this flux to estimate the radius of the
H$\beta$ emitting region at the time of the \textit{Suzaku}
campaign. After calibration with a standard star, the optical
spectra were put on an absolute calibration scale assuming a
constant
 flux of $F$([OIII] 5007)$_{\textrm{standard}}$ = 5.58$\times$10$^{-13}$ ergs/s/cm$^2$
(Peterson et al. 1991) and found the monochromatic
 flux at 5100 {\AA}, $F(5100 {\textmd{
\AA}})_{\textmd{observed}}$ = 5.13$\times$10$^{-15}$
ergs/s/cm$^2/${\AA}. For the aperture size of $5\textmd{
arcsec}\times7.5\textmd{ arcsec}$, the contribution of the host
galaxy at 5100 {\AA} is
 4.47$\times$10$^{-15}$ ergs/s/cm$^2/${\AA} (Bentz et al. 2006).
 To account for the effect of the aperture and the difference between
telescopes, the flux measured by FLWO should be converted by the
coefficients given in Peterson et al. (2002), i.e. $F(5100 {\textmd{
\AA}})_{\textmd{true}}=\varphi F(5100 {\textmd{
\AA}})_{\textmd{observed}}-G$. We averaged the results using the
different coefficients (i.e. $\varphi$ and $G$) determined in years
8 and 9-13 of the campaign (see details in Peterson et al. 2002),
and then excluded the flux of the host galaxy
(4.47$\times$10$^{-15}$ ergs/s/cm$^2/${\AA}, Bentz et al. [2006]) .
The final 5100 {\AA} flux of the AGN in NGC 5548  is $(1.7\pm0.7)
\times 10^{-15}$ ergs/s/cm$^2/${\AA} during the \textit{Suzaku}
campaign. With the relation between the emitting region of H$\beta$
line and $F$(5100 {\AA}) (Figure 5 in Bentz et al. [2007]), we found
r(H$\beta$)= $8\pm2$ light days during the \textit{Suzaku}
observations.

 From the result of dust reverberation of NGC 5548,
we also know the inner radius of the hot dust in this source is
47-53 light days (from the lag of K-band relative to V-band,
Suganuma et al. 2006).

 We plot the radial locations of Fe K$\alpha$,
H$\beta$ and the inner dust in Figure 9 (note that this is a linear
plot). It is likely that the location of the Fe K emitting region is
closer in than the inner radius of the dust, but slightly farther
out than H$\beta$, i.e. the origin of the narrow iron line lies in
the outer part of the BLR [$(0.5-1.0)\times10^{4}$ $r_g$]. This
result is consistent with  the  region emitting  the optical Fe II
emission lines, which is less than several light weeks from the
continuum emitter (Vestergaard \& Peterson 2005). This region is
also consistent with the intermediate line region proposed by Zhu et
al. (2009) and the radius of the low ionization, low velocity
component of the warm absorber in NGC 5548 (r$<$3 pc, Krongold et
al. 2009). Despite the uncertainty of the column density of the Fe K
emitting region, it is surely much higher (see \S5.3) than that of
the warm absorber found in NGC 5548 ($10^{21}\sim10^{22}$ cm$^{-2}$,
Krongold et al. 2009). Therefore, It clearly does not lie along our
line of sight.

\subsection{Physical condition}\label{sec53n}
We can use the location derived above to estimate the density of the
emitting region from the ionization parameter and the distance.
Kallman et al. (2004) investigated the dependence of the line
profile of K$\alpha$ and K$\beta$ on the ionization parameter. The
peak energies of K$\alpha$ and K$\beta$ in NGC 5548 (see Figure 2a)
prefer a relatively low ionization parameter, i.e. $\log
\xi=-1\pm1$, where $\xi=L/nR^2$ $(\textmd{erg s$^{-1}$cm})$,
 $L$ is the Fe K ionizing luminosity of the continuum source (erg s$^{-1}$),
$n$ is the density of the emitting region (cm$^{-3}$), and $R$ is
the distance to the continuum source (cm). With $\log L=43.7\pm0.3$,
$\log \xi=-1\pm1$, and $R=30\pm10$ light days, we obtain $\log
n=10.9\pm1.1$, which is comparable with the density of the gas in
the BLR of NGC 5548 (Ferland et al. 1992; Goad \& Koratkar 1998;
Kaspi \& Netzer 1999).

\subsection{Theoretical intensity of the Fe line}\label{sec52}

The intensity and the equivalent width of the iron line can be
estimated theoretically (Krolik \& Kallman 1987; Yaqoob et al.
2001). We found the column density $N_H>10^{23}$ cm$^{-2}$ is
required to produce the observed intensity and the equivalent width
of the Fe K$\alpha$ line in our \textit{Suzaku} observations.
However, as pointed out by Miller et al.
 (2009) and Yaqoob et al. (2009), due to the self-absorption effect and the Compton scattering, for $N_H>10^{23}$ cm$^{-2}$, the relation between the
 intensity of
the Fe K line  and $N_H$ or the abundance of iron is quite
non-linear, and the
 intensity of
the Fe K line also
 significantly depends on the geometry of the emitting region and the observing
 angle. Since the detailed investigation of the geometry,
 $N_H$, and the abundance of the emitting region of Fe K lines is much beyond the scope of this
 paper, we will not further discuss the constraints obtained from
 the intensity and the equivalent width of the Fe K lines.

\section{Conclusions}\label{sec6}

We analyzed the iron K$\alpha$ and K$\beta$ lines in spectra of NGC
5548 obtained by \textit{Suzaku} XIS and summarize our results as
follows.

1. The iron K$\alpha$ line was well detected ($>6\sigma$) in all
seven observations and the K$\beta$ line was also detected
($>2\sigma$) in four observations (1, 2, 4, and 7).

2. Only a narrow iron line was found in the spectra. The line width
obtained by the added spectra is 38$_{-18}^{+16}$ eV, which is
consistent with the results of \textit{Chandra} and
\textit{XMM-Newton}. Assuming the same virial relation as that of
the  H$\beta$ line, this width corresponds to a radius of
20$_{-10}^{+50}$ light days. Any relativistically broadened disk
line must be a factor of 5 weaker than the narrow component in flux
at 90\% confidence level.

3. We compared the observed peak energies and intensity ratios of
K$\alpha$ and K$\beta$ lines  with the expected value and found they
are consistent with the low ionization states of iron, i.e. lower
than Fe XIII, at the 99\% confidence level.

4. The Fe K$\alpha$ line is consistent with being constant over the
50 days of the \textit{Suzaku} campaign, although the 3-10 keV
continuum varies by a factor of 4. It is shown that a location at
$>$100 light days is consistent with the data (see the discussion in
\S4.3 and Figure 6b), but this is not a unique result.

5. To further access the location of the iron lines using the light
curve, we calculated the transfer functions in spherical and disk
geometries, and compared the predicted light curves with the
observed one. The value of $\chi^2$ is smallest in the disk case
with $i=30^\circ$, which is better than the constant fitting at the
92.7\% level. The spherical thin shell case is also acceptable
(P=81\%). The inferred emitting radii are 27$_{-7}^{+22}$ light days
in the spherical case and 37$_{-5}^{+7}$ light days in the disk case
with $i=30^\circ$, which are
 consistent with that obtained from the width of iron lines.

 6. Combining the \textit{Suzaku} constraints, the
most likely origin of the narrow iron lines is about 20-40 light
days away from the central engine, i.e. the outer part of BLR
(5.2$\times10^{3}$-1.0$\times10^{4}$ $r_g$). However, we could not
completely rule out other possible origins.

The approaches  used in this paper offer a valuable tool for
determining the size and structure of the inner regions of AGNs,
 although we stress again this method is model dependent. The
constraint on the emitting region of the narrow Fe K line
 obtained by the width will be greatly improved by upcoming calorimeters, which have $>10$ time better
 the energy resolution, of only a few
eV.
 If future X-ray satellites (e.g. Astro-H, IXO, and Gen-X) with
larger effective area could reduce the error of the flux of the Fe K
emission line  by even a factor of 2, then it will be possible  to
 distinguish different geometries from the constant
flux fitting. Higher sampling frequency campaign, preferably over a
longer baseline, is also desirable to obtain a cross-correlation
function with a quality comparable to or better than current optical
observations.

 \acknowledgments{}
Y. L. Thanks Junfeng Wang and Joanna Kuraszkiewicz for the helpful
discussions. S.N.Z. acknowledges partial funding support by the
Yangtze Endowment from the Ministry of Education at Tsinghua
University, Directional Research Project  of the Chinese Academy of
Sciences under project no. KJCX2-YW-T03 and by the National Natural
Science Foundation of China under grant nos. 10821061, 10733010,
10725313, and by 973 Program of China under grant 2009CB824800. This
work has been funded by CXC grant GO7-8136A and Suzaku Grant
NNX08AB81G.

\begin{figure}
\begin{center}
\includegraphics[scale=.70]{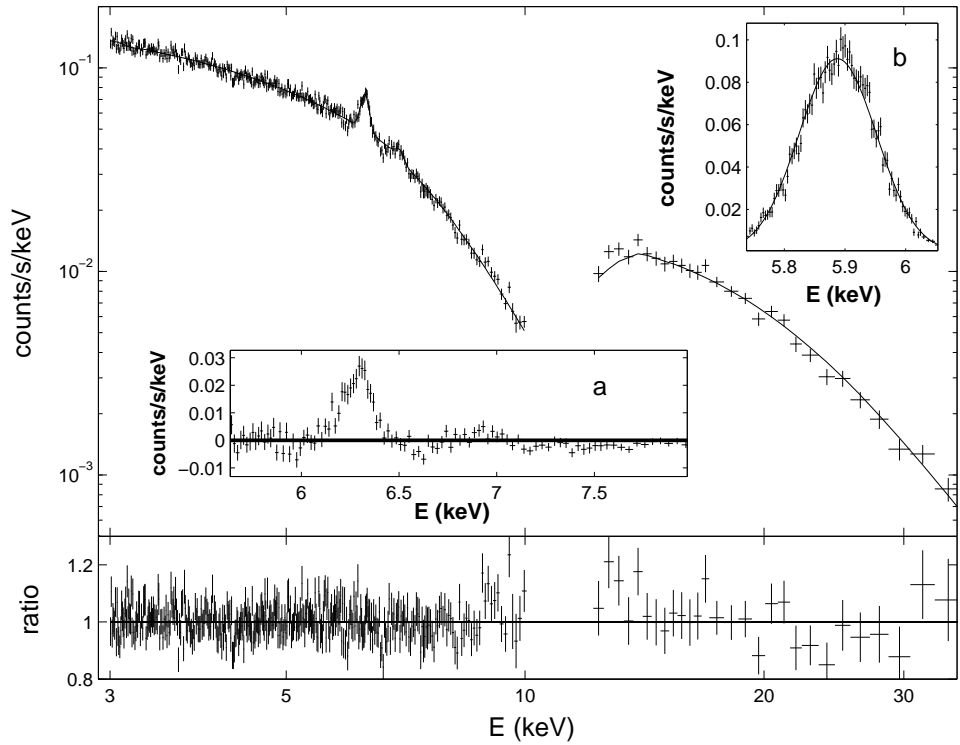}
\caption{\label{fig1}  The best fit model (solid lines) and the
added spectra of XIS03 (plus). The inset (a) is the residual of the
spectra of XIS03 to the best fit of a single power law.  The inset
(b) is the added spectra of Mn K$\alpha$ line of the calibration
sources on XIS03.}
\end{center}
\end{figure}

\begin{figure}
\begin{center}
\includegraphics[scale=.70]{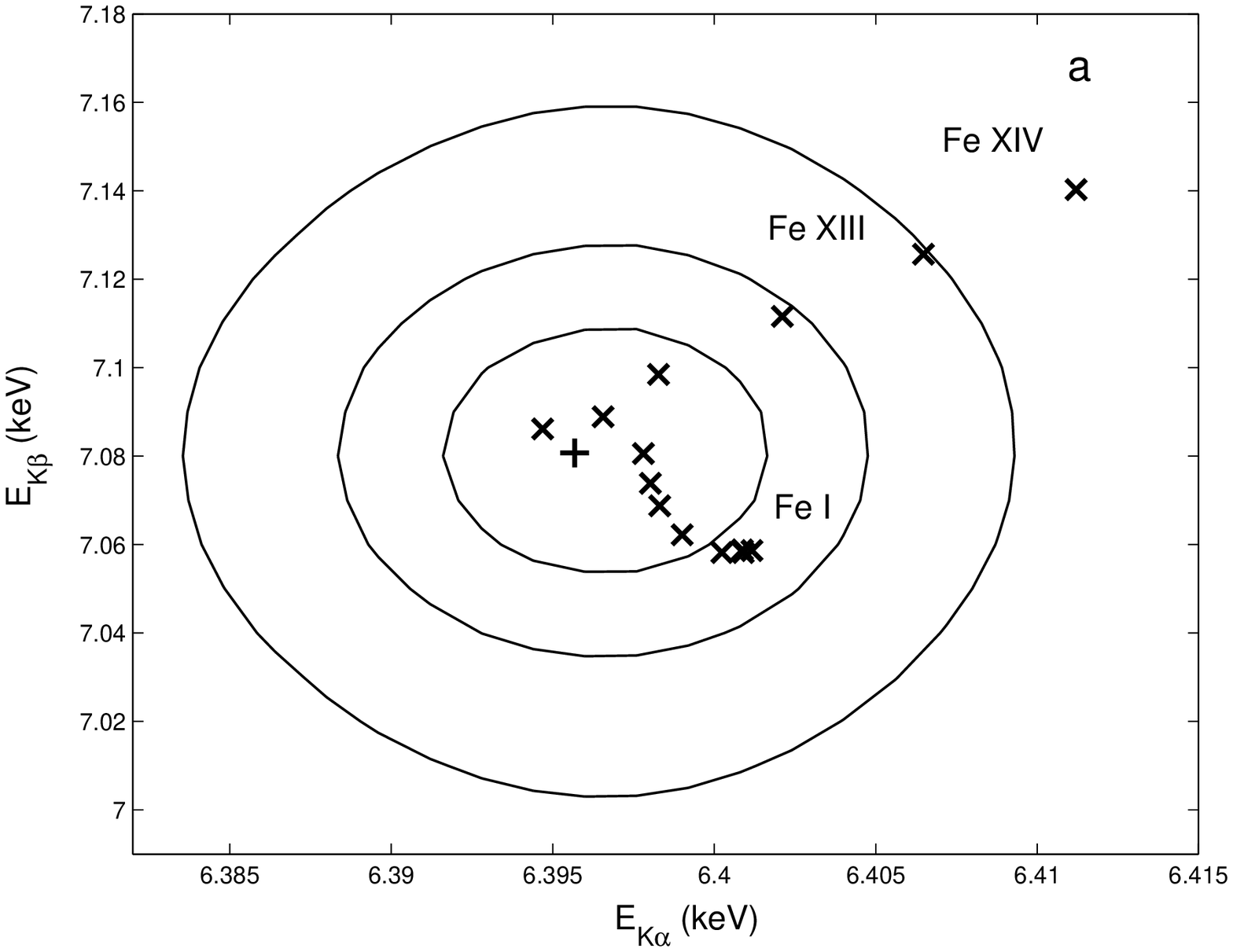}
\includegraphics[scale=.70]{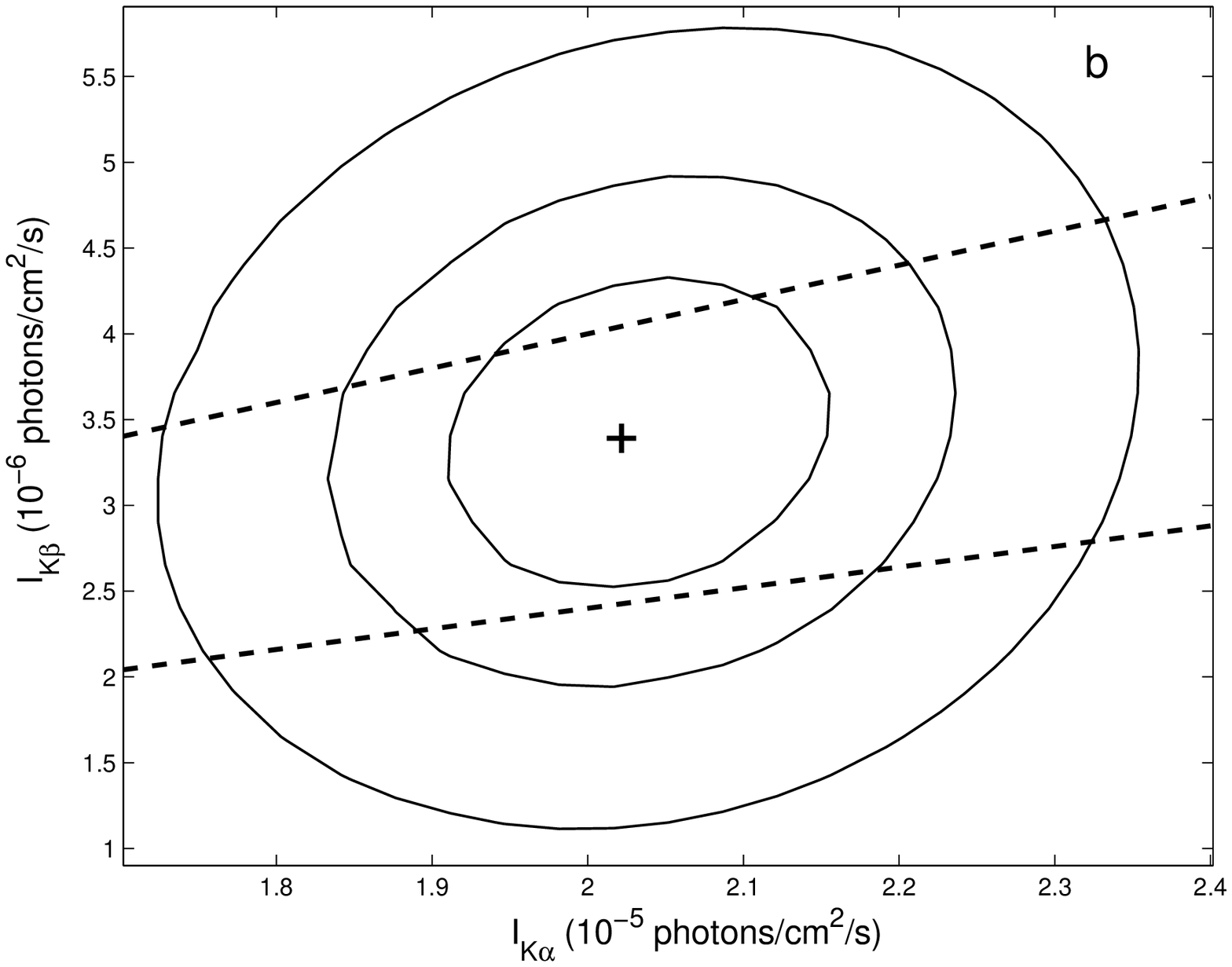}
\caption{\label{fig2} The confidence region of the peak energies (a)
and intensities (b) of K${\alpha}$ and K$\beta$ obtained from the
fitting of the co-added XIS03 and PIN spectra. The contours from
inner to outer correspond to $\Delta \chi^2$=1.00, 2.71, and 6.63
(68\%, 90\%, 99\%), respectively. The crosses in (a) are the
predicted values of the peak energies in different ionization states
of iron. The lower and upper dashed lines in (b) correspond to the
I(K$\beta$)/I(K${\alpha})$ ratios of 0.12 and 0.20, respectively. }
\end{center}
\end{figure}

\begin{figure}
\begin{center}
\includegraphics[scale=.70]{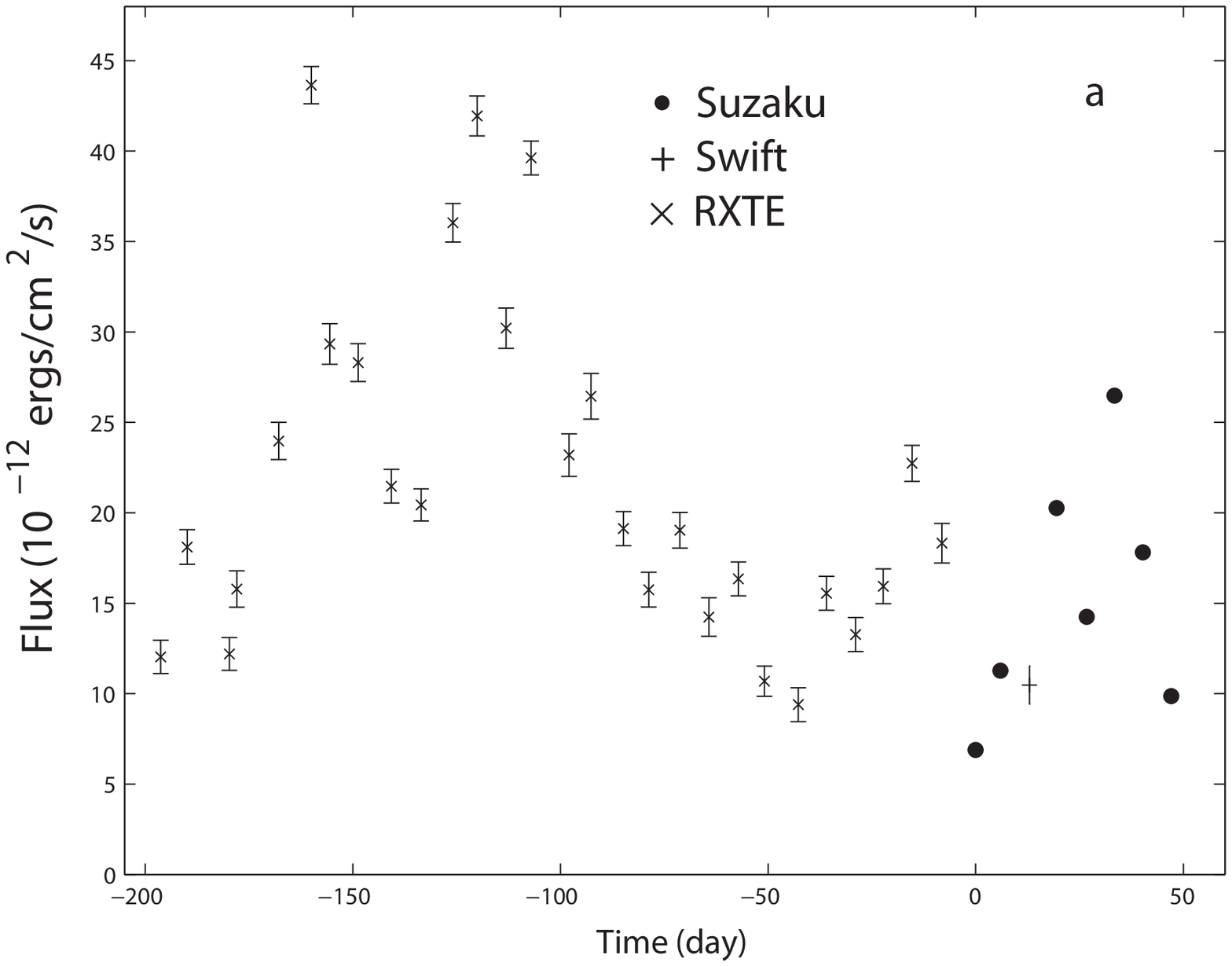}
\includegraphics[scale=.70]{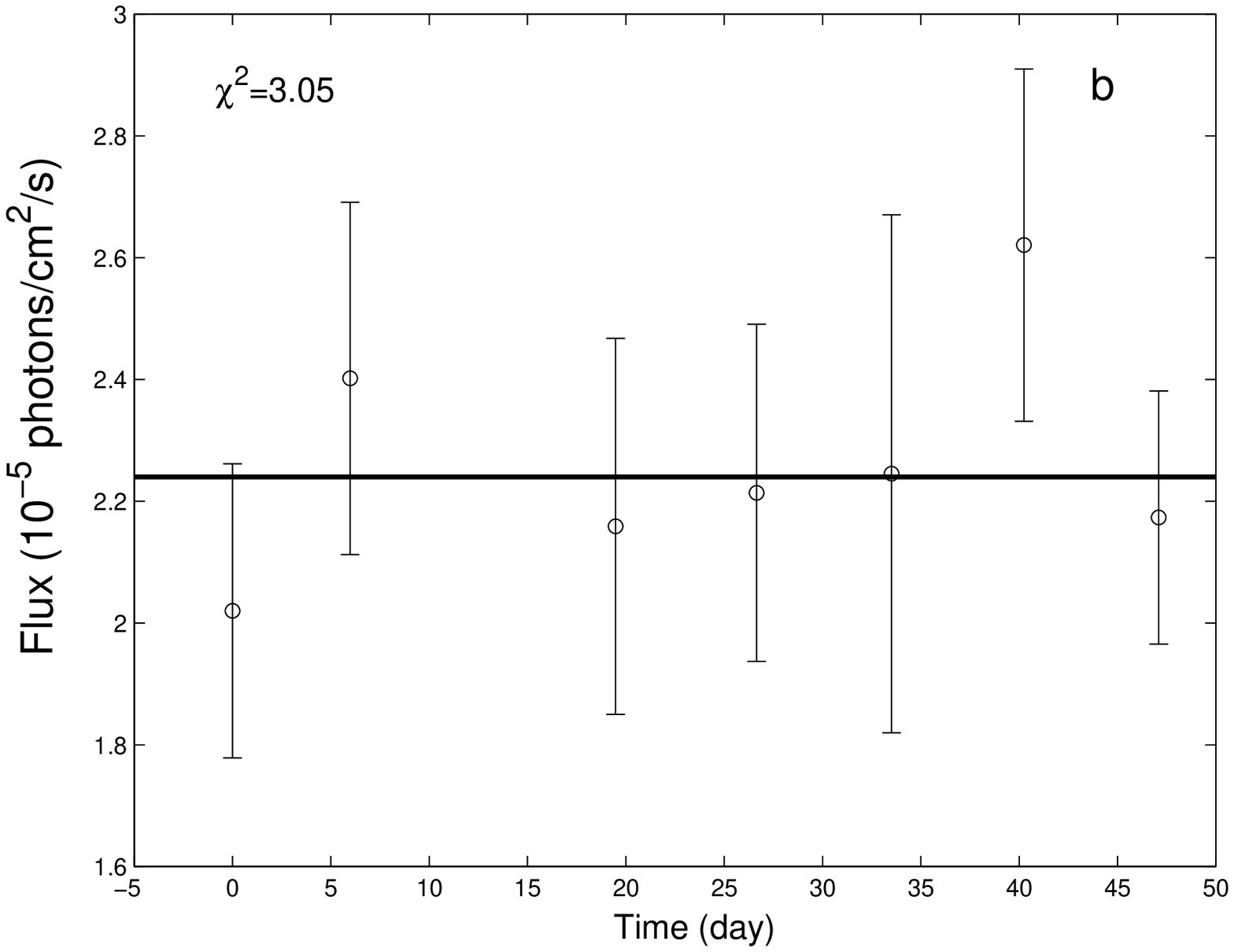}
\caption{\label{fig4} (a) The light curve of the flux of the
continuum in 3-10 keV band. The small error bar of the flux of the
\textit{Suzaku} observation is omitted. (b) The light curve of the
flux of K$\alpha$ line. The solid line in (b) is the result of the
constant fitting (see the text in \S4.1). }
\end{center}
\end{figure}

\begin{figure}
\begin{center}
\includegraphics[scale=.70]{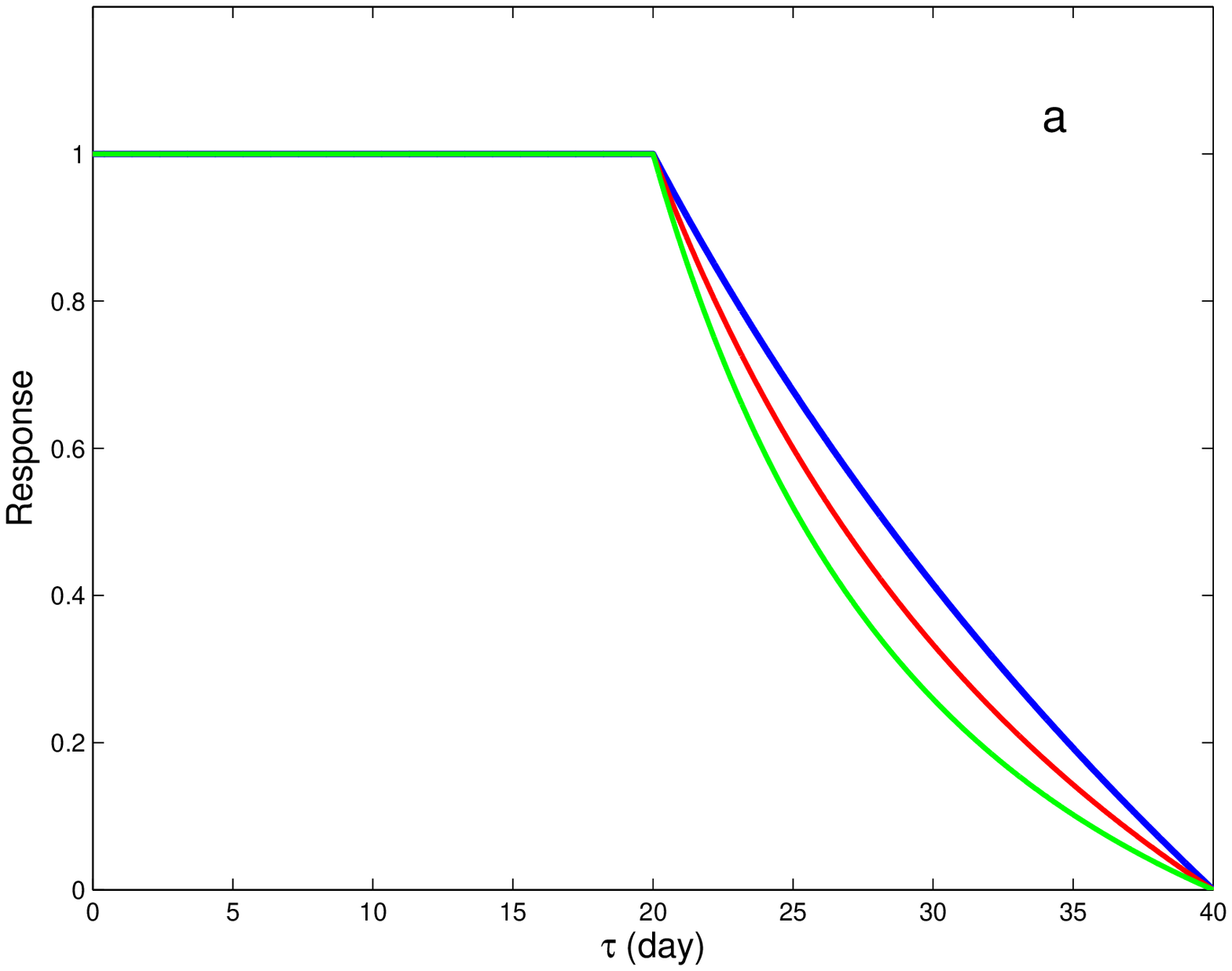}
\includegraphics[scale=.70]{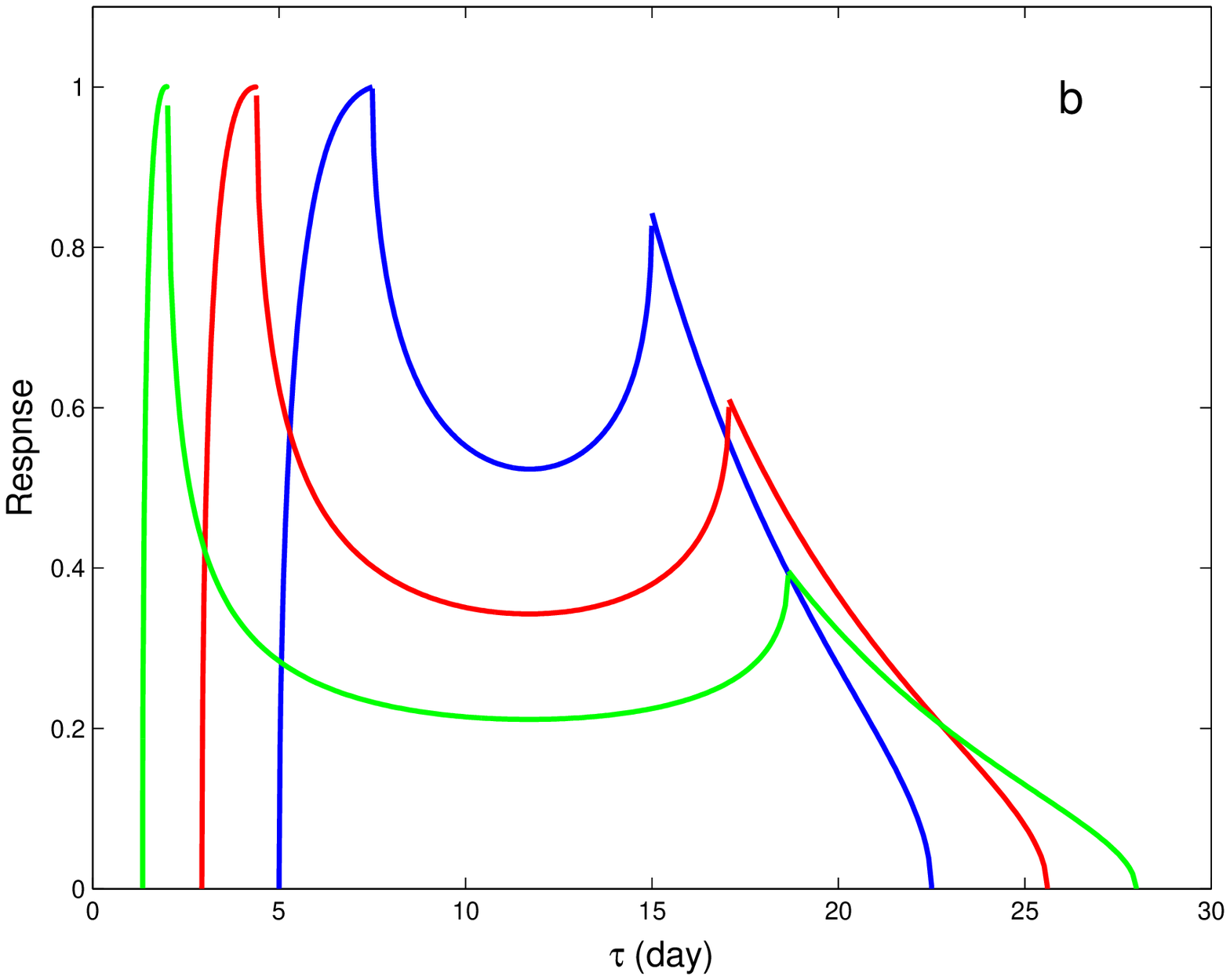}
\caption{\label{fig6} The transfer functions for the spherical (a)
and disk cases (b). We choose $r_{min}=10$ light days and
$r_{max}=20$ light days for all curves, which are normalized to 1 at
the maximum. (a) $\alpha=2$ (blue), $\alpha=3$ (red), and $\alpha=4$
(green). (b) $\alpha=3$ for all curves. $i=30^\circ$ (blue),
$i=45^\circ$ (red), and $i=60^\circ$ (green).}
\end{center}
\end{figure}

\begin{figure}
\begin{center}
\includegraphics[scale=.70]{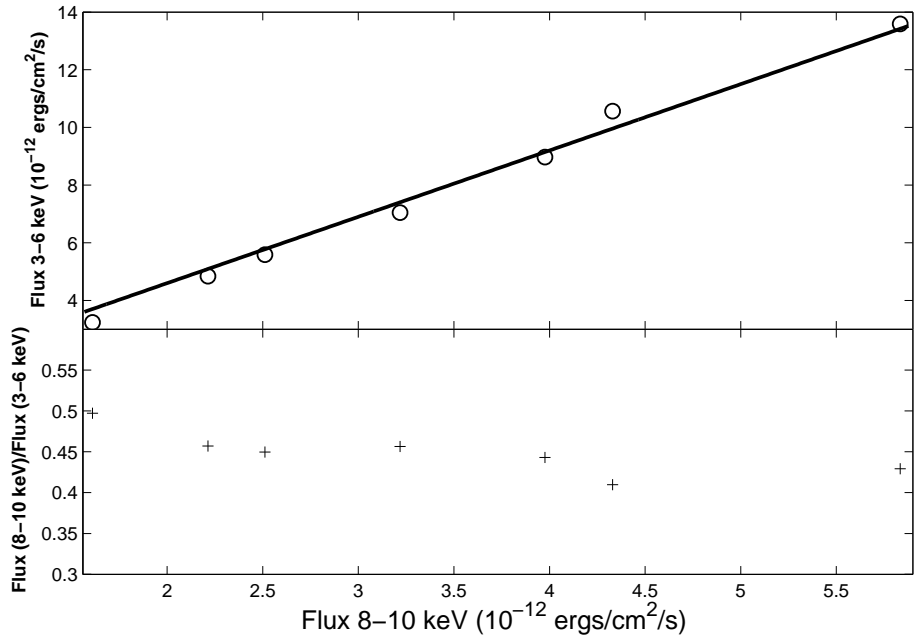}

\caption{\label{fig7n} The top panel shows the correlation between
the flux in 3-6 keV band and that in 8-10 keV band of all seven
observations. The solid line is the best-fitting straight line
across the origin. The error bars of the fluxes are omitted, since
they are smaller than the symbols. The bottom panel shows the ratios
between the flux in 8-10 keV band and that in 3-6 keV band.}
\end{center}
\end{figure}

\begin{figure}
\begin{center}
\includegraphics[scale=.70]{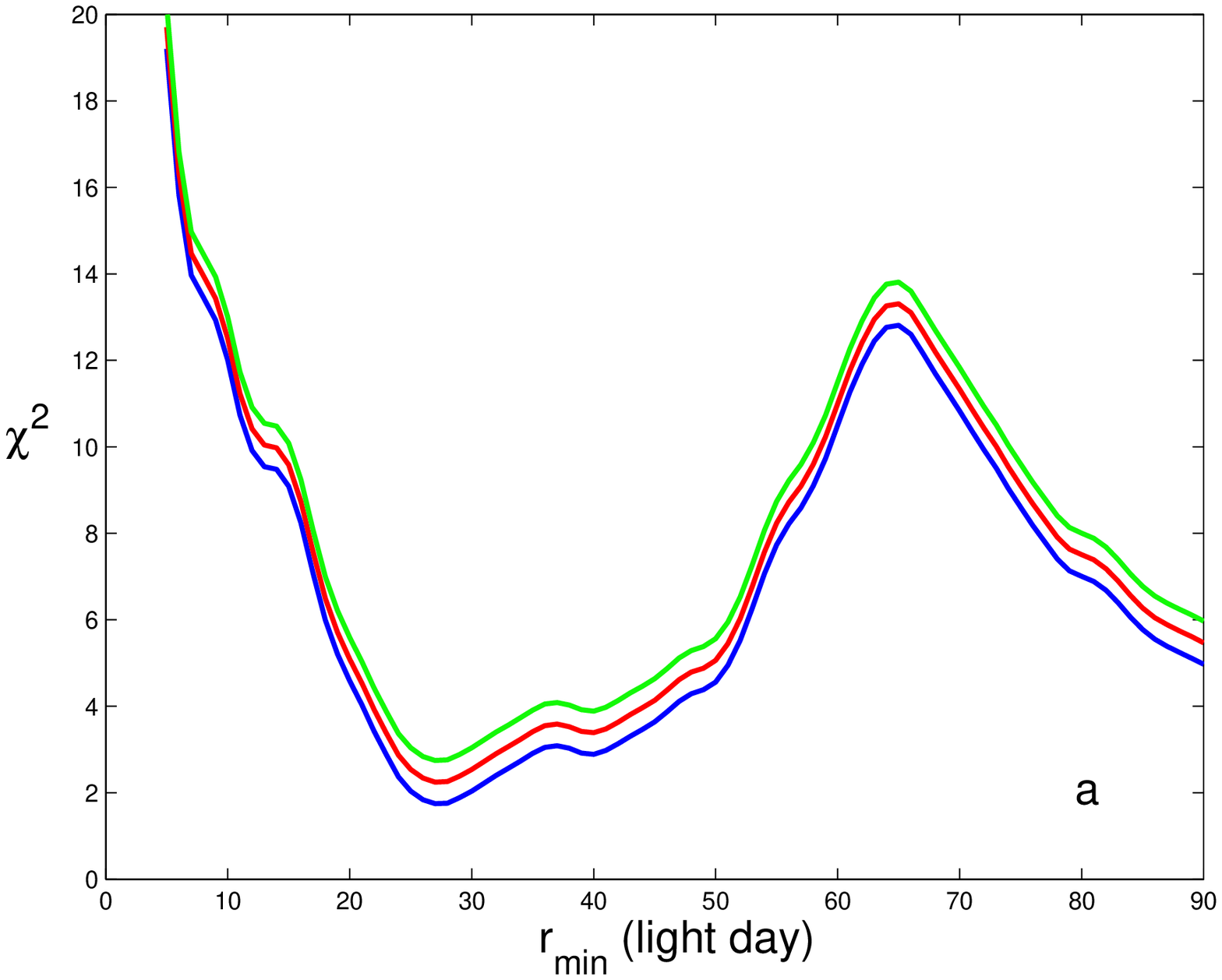}
\includegraphics[scale=.70]{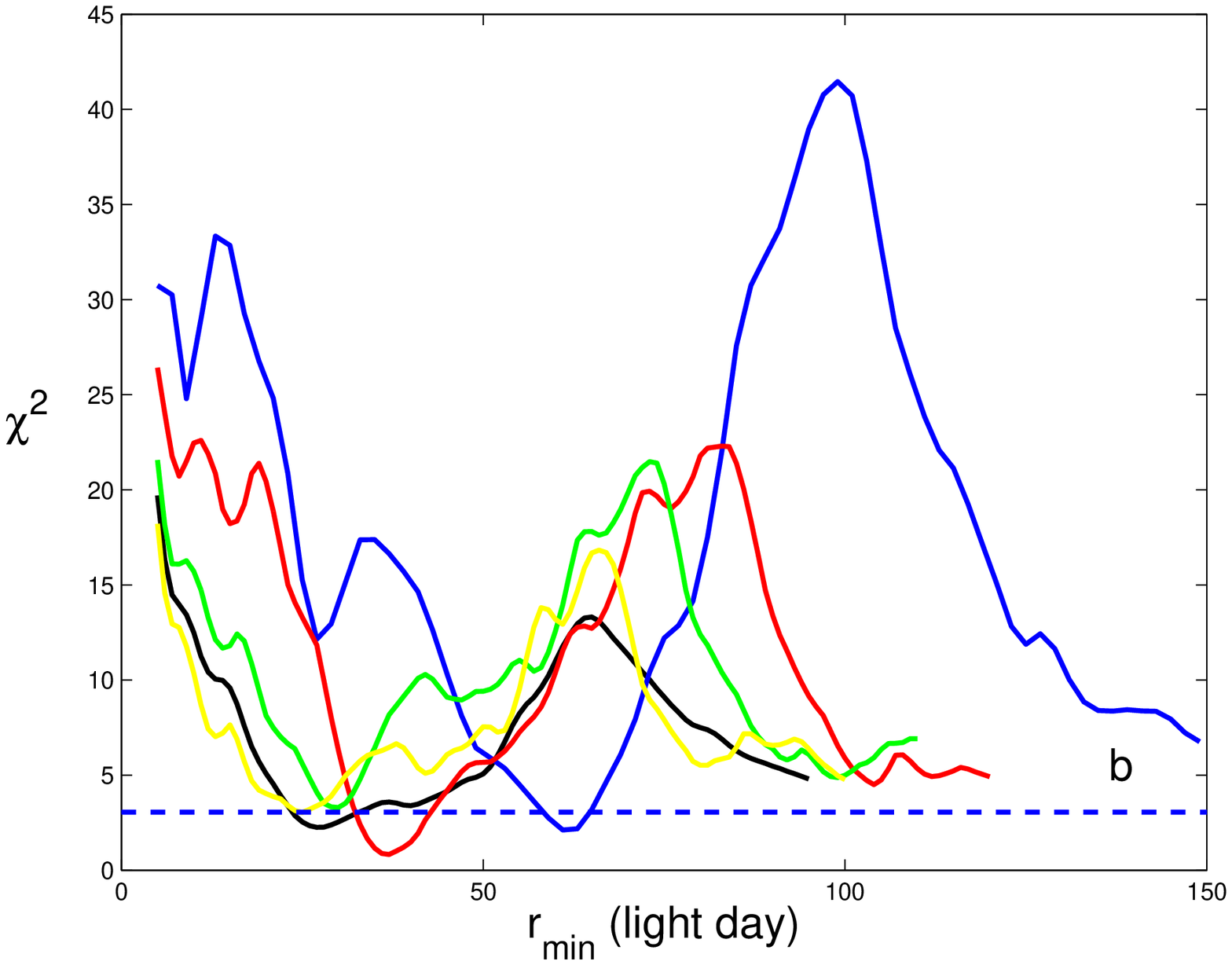}
\end{center}
\end{figure}

\begin{figure}
\begin{center}
\includegraphics[scale=.70]{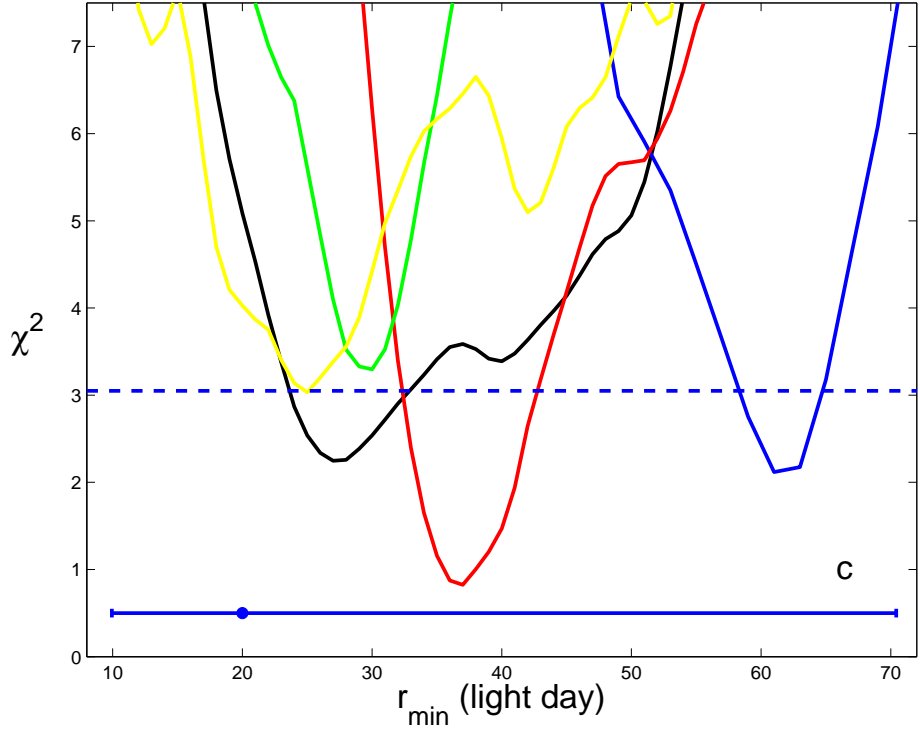}
\caption{\label{fig8} (a) Comparison with different values of
$\alpha$ in the spherical case. $\alpha=2$ (blue), $\alpha=3$ (red),
and $\alpha=4$ (green). Since the original curves are nearly
identical, the curves of $\alpha=2$ and $\alpha=4$ are vertically
shifted by -0.5 and 0.5, respectively. (b) The curves of $\chi^2$
with different $r_{min}$ for spherical and disk cases. Spherical
(black), disk $i=15^\circ$ (blue), $i=30^\circ$ (red), $i=45^\circ$
(green), and $i=60^\circ$ (yellow). The horizontal dashed line is
the $\chi^2$ of the constant fitting (see the text). (c) Enlarged
plot of (b). The error bar at the bottom is the 90\% confidence
interval inferred from the line width obtained by the simultaneous
fitting of all spectra (see Figure 8).}
\end{center}
\end{figure}

\begin{figure}
\begin{center}
\includegraphics[scale=.70]{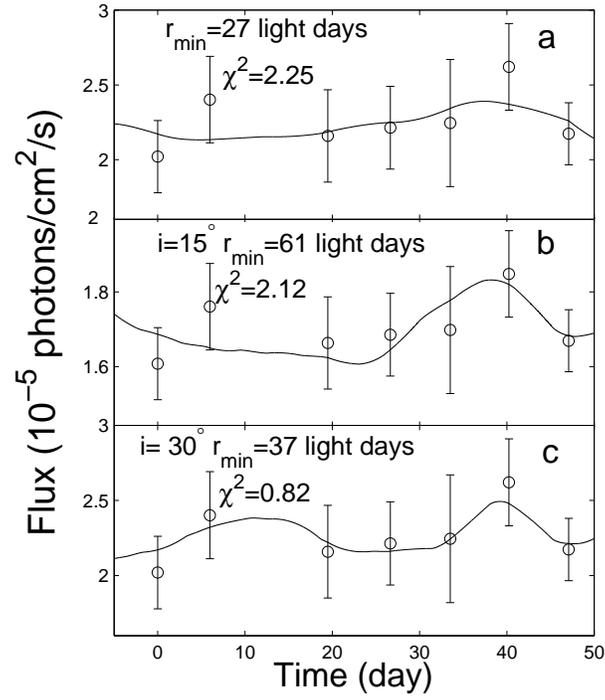}
\caption{\label{fig7} The comparison between the observed and
predicted light curve of the flux of the line with the minimum of
$\chi^2$ in different cases, i.e. (a) spherical case, (b) disk case
with $i=15^{\circ}$, (c) disk case with $i=30^{\circ}$.
}
\end{center}
\end{figure}

\begin{figure}
\begin{center}
\includegraphics[scale=.70]{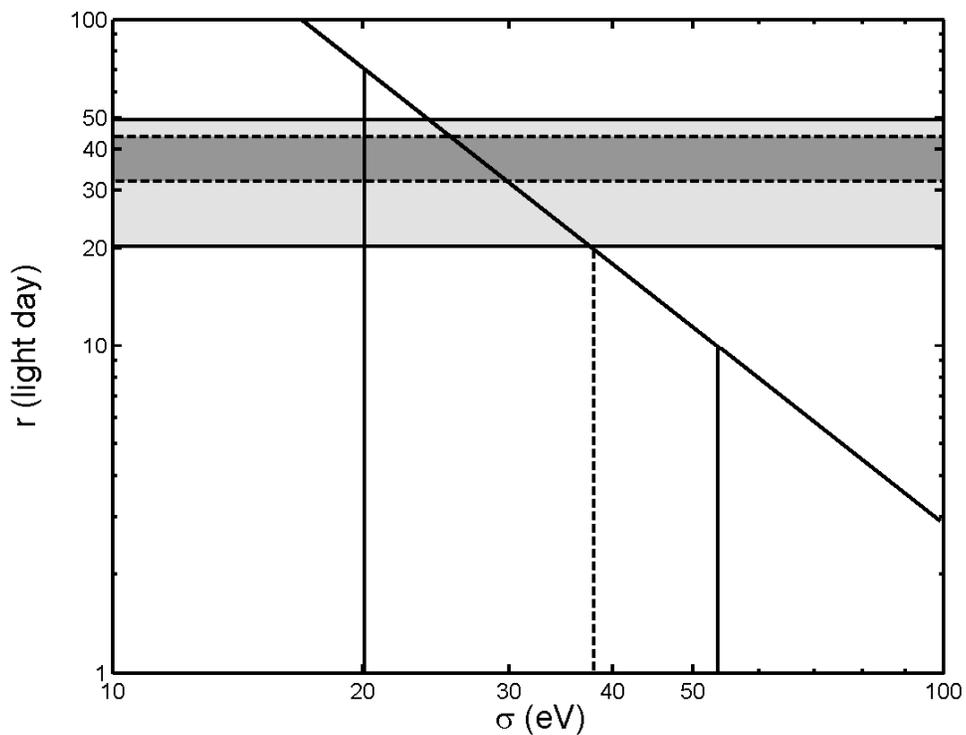}
\caption{\label{fig5} The relationship between $\sigma$ of the line
and the radius of the emitting region inferred from the black hole
mass. The vertical dashed and  solid lines correspond the central
value and the 90\% confidence intervals of the line width obtained
by the co-added XIS 03 and PIN spectra, respectively. The horizontal
  lines correspond to the 90\% confidence intervals of the
emitting region inferred from the light curve in the spherical case
(solid) and the disk case with $i=30^\circ$ (dashed, see Figure
6c).}
\end{center}
\end{figure}

\begin{figure}
\begin{center}
\includegraphics[scale=.70]{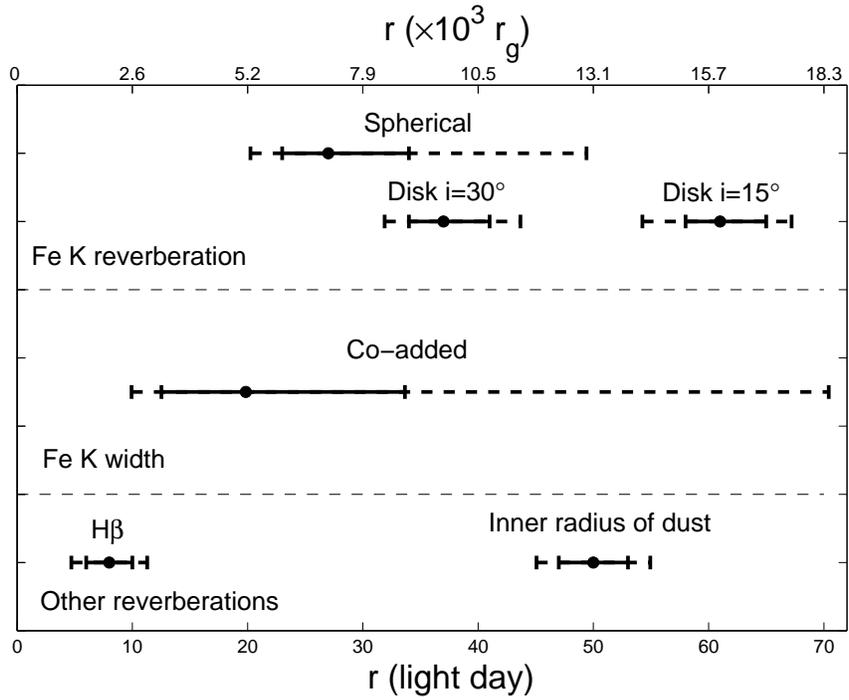}
\caption{\label{fig10} The summary of the location of the Fe
K$\alpha$ line emitting region. We show intervals derived from the
variation of the flux and the width of the Fe K line, with the
locations of H$\beta$ and the inner radius of the dust for
comparison. The solid and dashed error bars correspond to 68\% and
90\% confidence intervals, respectively.}
\end{center}
\end{figure}

\clearpage
\begin{table}[htp]\footnotesize
\begin{center}
\caption{\textit{Suzaku} observation log.\label{tbl1}}
\begin{tabular}{ccccc}
\tableline\tableline Sequence number & Obs. ID & Start Date \& Time
& Exposure time (ks) & 3-10 keV Count Rate
\\ & & & & ($10^{-3}$ photons/cm$^2$/s) \\
\tableline 1  & 702042010 &  2007-06-18 UT 22:28:15 & 31.1 & 0.772 \\
2 & 702042020 & 2007-06-24 UT 21:53:31 & 35.9 & 1.29 \\
3  & 702042040 &  2007-07-08 UT 10:02:55 & 30.7 & 2.35 \\
4 & 702042050 & 2007-07-15 UT 13:57:39 & 30.0 & 1.62 \\
5   & 702042060  & 2007-07-22 UT 10:40:25 & 28.9 & 3.05 \\
6   & 702042070 & 2007-07-29 UT 04:20:44 & 31.8 & 2.04 \\
7   & 702042080 & 2007-08-05 UT 00:37:46 & 38.8 & 1.12 \\

 \tableline
\end{tabular}
\end{center}
\end{table}
\clearpage
\begin{table}[htp]\footnotesize
\begin{center}
\caption{\textit{RXTE} and \textit{Swift} observation
logs.\label{tbl3}}
\begin{tabular}{cccc}
\tableline\tableline Obs. ID & Start Date \& Time & Exposure time
(s) & 3-10 keV Flux
\\ & &  & ($10^{-12}$ ergs/cm$^2$/s) \\
\tableline \textit{RXTE}& & &\\
\tableline
92113-07-40-00  &   2006-12-05  UT  03:35:48    &   919 &   12.0    $\pm$   0.9     \\
92113-07-41-00  &   2006-12-11  UT  12:59:06    &   885 &   18.1    $\pm$   1.0     \\
92113-07-42-00  &   2006-12-21  UT  16:34:27    &   961 &   12.2    $\pm$   0.9     \\
92113-07-43-00  &   2006-12-23  UT  11:08:04    &   830 &   15.8    $\pm$   1.0     \\
92113-07-44-00  &   2007-01-02  UT  13:09:22    &   914 &   24.0    $\pm$   1.0     \\
92113-07-45-00  &   2007-01-10  UT  10:09:27    &   1259    &   43.6    $\pm$   1.0     \\
92113-07-46-00  &   2007-01-14  UT  21:15:16    &   926 &   29.3    $\pm$   1.1     \\
92113-07-47-00  &   2007-01-21  UT  16:35:22    &   913 &   28.3    $\pm$   1.0     \\
92113-07-48-00  &   2007-01-29  UT  17:19:33    &   948 &   21.5    $\pm$   0.9     \\
92113-07-49-00  &   2007-02-05  UT  20:22:13    &   1171    &   20.4    $\pm$   0.9     \\
92113-07-50-00  &   2007-02-13  UT  12:12:30    &   895 &   36.0    $\pm$   1.1     \\
92113-07-51-00  &   2007-02-19  UT  08:05:31    &   896 &   41.9    $\pm$   1.1     \\
92113-07-52-00  &   2007-02-26  UT  07:25:07    &   1002    &   30.2    $\pm$   1.1     \\
92113-07-53-00  &   2007-03-04  UT  07:48:42    &   1354    &   39.6    $\pm$   0.9     \\
92113-07-54-00  &   2007-03-13  UT  11:30:51    &   614 &   23.2    $\pm$   1.2     \\
92113-07-55-00  &   2007-03-18  UT  18:56:43    &   572 &   26.4    $\pm$   1.3     \\
92113-07-56-00  &   2007-03-26  UT  13:58:42    &   994 &   19.1    $\pm$   0.9     \\
92113-07-57-00  &   2007-04-01  UT  16:10:48    &   982 &   15.7    $\pm$   1.0     \\
92113-07-58-00  &   2007-04-09  UT  04:12:03    &   893 &   19.0    $\pm$   1.0     \\
92113-07-59-00  &   2007-04-16  UT  03:23:00    &   955 &   14.2    $\pm$   1.1     \\
92113-07-60-00  &   2007-04-23  UT  06:05:57    &   961 &   16.3    $\pm$   0.9     \\
92113-07-61-00  &   2007-04-29  UT  12:54:16    &   1041    &   10.7    $\pm$   0.8     \\
92113-07-62-00  &   2007-05-07  UT  17:14:43    &   928 &   9.4     $\pm$   0.9     \\
92113-07-63-00  &   2007-05-14  UT  11:42:04    &   950 &   15.5    $\pm$   0.9     \\
92113-07-64-00  &   2007-05-21  UT  11:48:48    &   922 &   13.3    $\pm$   0.9     \\
92113-07-65-00  &   2007-05-28  UT  03:30:13    &   931 &   15.9    $\pm$   1.0     \\
92113-07-66-00  &   2007-06-04  UT  01:59:35    &   949 &   22.7    $\pm$   1.0     \\
92113-07-67-00  &   2007-06-11  UT  07:33:05    &   888 &   18.3    $\pm$   1.1     \\

\tableline \textit{Swift}& & &\\
\tableline
00030022061 & 2007-07-01 UT20:25:01 & 2503.18 & 10.5\\
 \tableline
\end{tabular}
\end{center}
\end{table}

\clearpage
\begin{landscape}
\begin{table}\footnotesize

\caption{The derived parameters from the spectral fits of the XIS
data of each observation (with error corresponding to 90\%
confidence level except flux and intensity).\label{tbl2}}
\begin{tabular}{c|cc|cccc|ccc|ccc}
\tableline Sequence & \multicolumn{2}{|c|}{Continuum} &
\multicolumn{4}{|c|}{K$\alpha$}& \multicolumn{3}{|c|}{K$\beta$}
\\
 number & $\Gamma$\tablenotemark{a}  & Flux\tablenotemark{b}  & E\tablenotemark{c} & $\sigma$\tablenotemark{d}  &I\tablenotemark{e}
  & EW\tablenotemark{f} & E\tablenotemark{c}   & I\tablenotemark{g}  & EW\tablenotemark{f}
  & $\chi^2$/dof& $\Delta\chi^2_1 \;\tablenotemark{h}$& $\Delta \chi^2_2\; \tablenotemark{i}$\\
\tableline 1&1.41$_{-0.06}^{+0.07}$&6.89$_{-0.34}^{+0.24}$&6.405$_{-0.017}^{+0.016}$&44$_{-44}^{+32}$&2.02$_{-0.24}^{+0.23}$&215$_{-77}^{+83}$&7.00$_{-0.08}^{+0.08}$&5.6$_{-1.8}^{+1.7}$&68$_{-68}^{+72}$&78.9/114&10.2&122.4\\
2&1.55$_{-0.05}^{+0.05}$&11.28$_{-0.30}^{+0.23}$&6.384$_{-0.016}^{+0.017}$&46$_{-46}^{+29}$&2.40$_{-0.29}^{+0.25}$&155$_{-56}^{+58}$&7.08$_{-0.06}^{+0.06}$&5.5$_{-2.0}^{+2.0}$&42$_{-42}^{+51}$&197.7/177&7.4&122.8\\
3&1.68$_{-0.03}^{+0.03}$&20.25$_{-0.16}^{+0.13}$&6.374$_{-0.025}^{+0.027}$&62$_{-37}^{+31}$&2.16$_{-0.31}^{+0.30}$&77$_{-38}^{+39}$&7.04\tablenotemark{j}&$<$5.5&$<$40&355.1/338&бн&65.7\\
4&1.53$_{-0.04}^{+0.04}$&14.24$_{-0.27}^{+0.24}$&6.380$_{-0.018}^{+0.019}$&34$_{-34}^{+31}$&2.21$_{-0.27}^{+0.28}$&112$_{-48}^{+48}$&7.05$_{-0.04}^{+0.04}$&7.5$_{-2.1}^{+2.2}$&44$_{-44}^{+44}$&227.3/214&11.7&98.9\\
5&1.61$_{-0.03}^{+0.03}$&26.46$_{-0.16}^{+0.14}$&6.401$_{-0.031}^{+0.029}$&62$_{-62}^{+49}$&2.25$_{-0.36}^{+0.43}$&61$_{-34}^{+34}$&7.04\tablenotemark{j}&$<$5.3&$<$31&427.9/406&бн&54.6\\
6&1.57$_{-0.03}^{+0.03}$&17.80$_{-0.22}^{+0.18}$&6.393$_{-0.015}^{+0.015}$&34$_{-34}^{+29}$&2.62$_{-0.26}^{+0.29}$&106$_{-39}^{+39}$&7.04\tablenotemark{j}&$<$7.0&$<$50&303.2/308&бн&124.1\\
7&1.53$_{-0.04}^{+0.04}$&9.85$_{-0.24}^{+0.20}$&6.416$_{-0.012}^{+0.012}$&23$_{-23}^{+28}$&2.17$_{-0.21}^{+0.19}$&162$_{-53}^{+52}$&6.97$_{-0.08}^{+0.09}$&4.0$_{-1.6}^{+1.5}$&34$_{-34}^{+43}$&205.0/216&6.9&166.1\\

 \tableline
\end{tabular}
\tablenotetext{a}{Photon index} \tablenotetext{b}{Observed flux in
3-10 keV band (10$^{-12}$ ergs/cm$^2$/s) with 68\% error}
\tablenotetext{c}{Rest energy of line (keV)}
\tablenotetext{d}{Intrinsic width of line (eV)}
\tablenotetext{e}{Observed intensity of line (10$^{-5}$
photons/cm$^2$/s) with 68\% error} \tablenotetext{f}{Equivalent
width of line (eV). 90\% upper limit is shown if the K$\beta$ line
is very weak.} \tablenotetext{g}{Observed intensity of line
(10$^{-6}$ photons/cm$^2$/s) with 68\% error. 90\% upper limit is
shown if the K$\beta$ line is very weak.}\tablenotetext{h}{The
increase of $\chi^2$ after removing K$\beta$ line from the best-fit
model}\tablenotetext{i}{The increase of $\chi^2$ after removing
K$\alpha$ and K$\beta$ lines from the best-fit model}
\tablenotetext{j}{The rest energy of line is fixed at the value
obtained by the co-add spectra, since the K$\beta$ line is very weak
in this observation}
\end{table}
\end{landscape}

\clearpage
\begin{landscape}
\begin{table}\footnotesize

\caption{The derived parameters from the spectral fits of the
co-added XIS03 and PIN data (with error corresponding to 90\%
confidence level except flux and intensity).\label{tbl2}}
\begin{tabular}{c|cccc|cccc|ccc|ccc}
\tableline Sequence &
\multicolumn{4}{|c|}{Continuum\tablenotemark{a}} &
\multicolumn{4}{|c|}{K$\alpha$}& \multicolumn{3}{|c|}{K$\beta$}
\\
 number & $\Gamma$\tablenotemark{b}  & E$_{\textrm{cut-off}}$\tablenotemark{c} & R\tablenotemark{d} & Flux\tablenotemark{b}  & E\tablenotemark{b} & $\sigma$\tablenotemark{b}  &I\tablenotemark{b}
  & EW\tablenotemark{b} & E\tablenotemark{b}   & I\tablenotemark{b}  & EW\tablenotemark{b}
  & $\chi^2$/dof\tablenotemark{b}& $\Delta\chi^2_1 \;$\tablenotemark{b}& $\Delta \chi^2_2\; $\tablenotemark{b}\\
\tableline 1&1.59$_{-0.03}^{+0.03}$& 75$_{-15}^{+85}$ & 0.79
$_{-0.32}^{+0.35}$&15.1$_{-0.3}^{+0.1}$
&6.396$_{-0.007}^{+0.009}$&38$_{-18}^{+16}$&2.02$_{-0.11}^{+0.13}$&95$_{-19}^{+21}$&7.08$_{-0.05}^{+0.05}$
&3.4$_{-0.9}^{+0.9}$&19$_{-17}^{+17}$&423.8/396&14.7&405.5\\

 \tableline
\end{tabular}
\tablenotetext{a}{The cosine of the disc inclination was fixed at
0.87} \tablenotetext{b}{The meaning is the same as that in Table
3}\tablenotetext{c}{Energy cut-off (keV)}
\tablenotetext{d}{Reflection parameter of cold matter;
$R=\Omega/2\pi$}
\end{table}
\end{landscape}

\end{document}